\documentclass[twocolumn,apl,amsmath,amssymb,showpacs,superscriptaddress]{revtex4-1}
\usepackage[utf8]{inputenc}
\usepackage{epsf}      
\usepackage{graphicx}
\usepackage{color}
\usepackage{gensymb}
\usepackage{sidecap}

\begin{document}

\title {Electronic structure of {\it A}Co$_2$As$_2$ ({\it A} = Ca, Sr, Ba, Eu) studied using angle-resolved photoemission spectroscopy and theoretical calculations}

\author{R. S. Dhaka}
\email{Corresponding author: rsdhaka@physics.iitd.ac.in}
\affiliation{The Ames Laboratory, U.S. DOE and Department of Physics and Astronomy, Iowa State University, Ames, Iowa 50011, USA}
\affiliation {Department of Physics, Indian Institute of Technology Delhi, Hauz Khas, New Delhi-110016, India}
\author{Y. Lee}
\affiliation{The Ames Laboratory, U.S. DOE and Department of Physics and Astronomy, Iowa State University, Ames, Iowa 50011, USA}
\author{V. K. Anand}
\email{Present address: Department of Physics, University of Petroleum and Energy Studies, Dehradun, Uttarakhand, 248007, India}
\affiliation{The Ames Laboratory, U.S. DOE and Department of Physics and Astronomy, Iowa State University, Ames, Iowa 50011, USA}
\author{Abhishek Pandey}
\email{Present address: Materials Physics Research Institute, School of Physics, University of the Witwatersrand, Johannesburg, Gauteng 2050, South Africa}
\affiliation{The Ames Laboratory, U.S. DOE and Department of Physics and Astronomy, Iowa State University, Ames, Iowa 50011, USA}
\author{D. C. Johnston}
\affiliation{The Ames Laboratory, U.S. DOE and Department of Physics and Astronomy, Iowa State University, Ames, Iowa 50011, USA}
\author{B. N. Harmon}
\affiliation{The Ames Laboratory, U.S. DOE and Department of Physics and Astronomy, Iowa State University, Ames, Iowa 50011, USA}
\author{Adam Kaminski}
\affiliation{The Ames Laboratory, U.S. DOE and Department of Physics and Astronomy, Iowa State University, Ames, Iowa 50011, USA}

\date{\today}                                    

\begin{abstract}
We present a comprehensive study of the low-energy band structure  and Fermi surface (FS) topology of {\it A}Co$_2$As$_2$ ({\it A} = Ca, Sr, Ba, Eu) using high-resolution angle-resolved photoemission spectroscopy. The experimental FS topology and band dispersion data are compared with theoretical full-potential linearized augmented-plane-wave (FP-LAPW) calculations, which yielded reasonably good agreement. We demonstrate that the FS maps of $A$Co$_2$As$_2$ are significantly different from those of the parent compounds of Fe-based high-temperature superconductors. Further, the FSs of CaCo$_2$As$_2$ do not show significant changes across its antiferromagnetic transition temperature. The band dispersions extracted in different momentum $(k_{\it x}, k_{\it y})$ directions show a small electron pocket at the center and a large electron pocket at the corner of the Brillouin zone (BZ). The absence of the hole FS in these compounds does not allow nesting between pockets at the Fermi energy ({\it E}$_{\rm F}$), which is in contrast to {\it A}Fe$_2$As$_2$-type parent compounds of the iron-based superconductors. Interestingly, we find that the hole bands are moved 300--400~meV below $E_{\rm F}$ depending on the $A$ element. Moreover, the existence of nearly flat bands in the vicinity of $E_{\rm F}$ are consistent with the large density of states at {\it E}$_{\rm F}$. These results are important to understand the physical properties as well as the possibility of the emergence of superconductivity in related materials.
\end{abstract}

\maketitle

\section{Introduction}

Understanding the origin of the unconventional pairing mechanism in high-temperature superconductors, cuprates \cite{Bednorz86, Schilling93, Chu93} and iron pnictides \cite{Kamihara08, Takahashi08, XHChen08, GFChen08, Rotter08, Yuan09, Cruz08, SefatPRL08, WangSc11}, is one of the most challenging issues in condensed-matter physics. In particular, the superconductivity (SC) in the iron pnictides has received enormous attention to find its intimate connection with the magnetic ordering \cite{Johnston10, CanfieldRev10}. For example, the parent compound  BaFe$_2$As$_2$ shows a phase change from high temperature tetragonal to low temperature orthorhombic structure at $\approx135$~K along with an associated antiferromagnetic (AFM) spin density wave (SDW) transition. It is well known that the common method to induce high-{\it T}$_{\rm c}$ superconductivity in these materials is to suppress the SDW ground state by different ways such as partial chemical substitution and/or application of external pressure \cite{Takahashi08, XHChen08, GFChen08, Rotter08, Yuan09, Cruz08, SefatPRL08, Johnston10, CanfieldRev10}. One of the most intriguing aspects is to study how the long-range magnetic order and superconductivity are related in these materials \cite{Cruz08, PrattPRL09}. In this context, investigation of the electronic structure and Fermi surface (FS) topology using angle-resolved photoemission spectroscopy (ARPES) is vital \cite{LiuPRB09, YangPRL09, ZhangPRL09, KondoPRB10, LiuPRL09, YiPRB09, DhakaPRL11, LiuNP10}. The FS of BaFe$_2$As$_2$ consists of a hole (at the $\Gamma$ point) and an electron (at the $X$ point) pockets, and their similar volumes suggest that the material is compensated since the number of electrons and holes are equal. The nesting between the hole and the electron FSs can give SDW ordering \cite{SinghPRL08, Mazin10, MazinPRL08} and play a pivotal role in driving the antiferromagnetic-paramagnetic phase transition \cite{Cruz08, Mazin10, DhakaPRL13}. The FSs of these materials show a remarkable reconstruction at low temperature due to the presence of an AFM SDW phase \cite{LiuPRB09, YangPRL09, ZhangPRL09, KondoPRB10, LiuPRL09, YiPRB09}, and the suppression of magnetic ordering is linked to the onset of the SC dome \cite{DhakaPRL11, LiuNP10}. Interestingly, when Fe is replaced by Co in BaFe$_2$As$_2$, a rigid-band-like change in the band structure occurs and Lifshitz transitions are observed both at the onset and offset of the SC dome \cite{LiuNP10, LiuPRB11, Thirupathaiah, Brouet, Sekiba, Ideta}. On the other hand, no significant changes in the FS and band structure were observed for isoelectronic Ru substituted BaFe$_2$As$_2$ across the SC dome \cite{DhakaPRL11, LiuPRB15}. 

Therefore, one approach is to completely substitute Co/Ru at the Fe site and investigate how the band structure and FS topology are related to the magnetic ordering as well as the SC in these materials. In this direction, very interesting structural and magnetic properties of {\it A}Co$_2$As$_2$,({\it A} = Ca, Sr, Ba, Eu) compounds have been reported \cite{VivekPRB14, BingPRB19, SangeethaPRB18, SapkotaPRL17,  DingPRB17, JayasekaraPRB15, Ying12, Cheng12, Bishop10, Sefat09, Pandey13}. For example, powder x-ray diffraction and magnetization measurements on CaCo$_2$As$_2$ demonstrate the structure to be collapsed-tetragonal and A-type collinear AFM order is observed below {\it T}$_{\rm N}=$ 52~K, respectively \cite{VivekPRB14}. However, the resistivity and specific heat measurements show no evidence of the magnetic transition in CaCo$_2$As$_2$ \cite{VivekPRB14}. On the other hand, BaCo$_2$As$_2$ exhibits paramagnetic behavior and no magnetic ordering is reported down to 1.8~K \cite{Sefat09, AnandPRB14}. Interestingly, these materials crystallize in the ThCr$_2$Si$_2$-type tetragonal structure and the $c/a$ ratio of SrCo$_2$As$_2$ is found to be 2.99, which is intermediate to those of normal-tetragonal BaCo$_2$As$_2$ ($c/a=$ 3.20) exhibiting ferromagnetic correlations and collapsed-tetragonal-antiferromagnetic CaCo$_2$As$_2$ ($c/a=$ 2.58) \cite{Pandey13}. Furthermore, the magnetization, NMR, and neutron diffraction measurements of SrCo$_2$As$_2$ show no evidence for long-range magnetic ordering above 0.05~K \cite{Pandey13, Jasper08, Jayasekara13, BingPRB19}. This is consistent with the isostructural BaCo$_2$As$_2$ \cite{Sefat09}, but is in contrast to the AFM behavior of CaCo$_2$As$_2$ \cite{VivekPRB14, Ying12, Cheng12}. More interestingly, SrCo$_2$As$_2$ shows a negative (positive) thermal expansion coefficient along the $c$ ($a$)-axis in the temperature range from $\approx$7~K to 300~K \cite{Pandey13}. Also, the structural properties of EuCo$_2$As$_2$ indicate a phase transition from tetragonal at ambient pressure to collapsed-tetragonal at high pressure \cite{Bishop10}. The magnetization measurements at ambient pressure indicate that the effective paramagnetic moment at high temperatures arises mainly from the Eu spins with an appearance of AFM ordering $\le 39$~K \cite{Ballinger12}.  

Notably, there have been many reports on the structural and magnetic properties of these {\it A}Co$_2$As$_2$ compounds \cite{VivekPRB14, BingPRB19, SangeethaPRB18, SapkotaPRL17,  DingPRB17, JayasekaraPRB15, Ying12, Cheng12, Bishop10, Sefat09, Pandey13}; however, detailed ARPES studies are very few \cite{DhakaPRB13, XuPRX13, MansartPRB16} and not reported for EuCo$_2$As$_2$. Therefore, further investigations of the band structure and FSs of {\it A}Co$_2$As$_2$ and comparison with the {\it A}Fe$_2$As$_2$ parent compounds are desired to shed light on the pairing mechanism of Fe-based superconductors. 

In this paper, we report a comprehensive ARPES study of {\it A}Co$_2$As$_2$ ({\it A} = Ca, Sr, Ba, Eu) and present the FSs as well as low-energy band structures, which are found to be very different from BaFe$_2$As$_2$ at the Fermi level. The experimental FSs and band dispersion data are compared with full-potential linearized augmented-plane-wave (FP-LAPW) calculations, which are found to be in reasonably good agreement. The corresponding band dispersion data show a small electron pocket at the center and large electron pocket at the corner of the Brillouin zone (BZ). This reveals that no obvious FS nesting is present in these compounds, which is in contrast to the parent compounds of Fe-based high-{\it T}$_{\rm c}$ superconductors. Moreover, we observe that the hole bands are moved 300--400~meV below the {\it E}$_{\rm F}$ compared to the hole bands crossing {\it E}$_{\rm F}$ in the {\it A}Fe$_2$As$_2$ compounds. Here in case of {\it A}Co$_2$As$_2$, the rigid-band-like shift in the band structure results in the appearance of a small electron pocket at the $\Gamma$ point \cite{DhakaPRB13, XuPRX13, MansartPRB16}. Moreover, we find that the bands in the vicinity of {\it E}$_{\rm F}$ are reasonably flat, which are responsible for high intensity peak in the density of states (DOS) at {\it E}$_{\rm F}$ and important to understand the physical properties of these $A$Co$_2$As$_2$ materials. In addition, there are no notable changes in the FS topology of CaCo$_2$As$_2$ measured at, below, and above the AFM transition temperature. 

\section{Experimental details}

High-quality single crystals of {\it A}Co$_2$As$_2$ ({\it A} = Ca, Sr, Eu) were grown with Sn-flux whereas BaCo$_2$As$_2$ out of self flux and the details of bulk physical property measurements and x-ray analysis can be found in Refs.~\cite{Pandey13,VivekPRB14, AnandPRB14, SangeethaPRB18}. We used a Scienta R4000 electron analyzer to perform high-resolution ARPES measurements at beamline 7.0.1 of the Advanced Light Source (ALS), Berkeley, California. All samples were cleaved {\it in situ} in an ultrahigh vacuum chamber having $\le$$4\times10^{-11}$~mbar pressure, yielding flat mirror-like surfaces in the {\it ab} plane. The ARPES data were collected with $\sim20$~meV and $\sim0.3^{\circ}$ energy and momentum resolutions, respectively. Several samples from different batches were measured to reproduce the results of the Fermi surfaces and band structure. We use a gold sample to determine the Fermi energy. 

We calculate the Fermi surfaces and band dispersions of $A$Co$_2$As$_2$ with the FP-LAPW method using the local density approximation \cite{Perdew92}. We use $R_{\rm MT}\times{\it k}_{\rm max}=$ 8 or 9 to find the self-consistent charge density, where the smallest muffin tin (MT) radius is multiplied by the maximum $k$ value in the plane wave expansion basis. Here, the MT radii of 2.1, 2.3, 2.5, 2.5, 2.1, and 2.1~a.u. were taken for Ca, Sr, Ba, Eu, Co, and As, respectively. Note that 828~$k$-points were selected in the irreducible BZ to perform the calculations untill we reached the total energy convergence criterion of 0.01 mRy/ primitive cell. Also, we relaxed the As atom {\it z}-axis position to find a minimum total energy, which gave {\it z}$_{\rm As}=$ 0.3622, 0.3515, 0.3441, and 0.3611 for $A=$ Ca, Sr, Ba, and Eu, respectively. To compute the FSs, we divide the $-2\pi/a \le (k_{\it x}$, $k_{\it y})\le 2\pi/a$ ranges of $k_{\it x}$, $k_{\it y}$ planes with different $k_{\it z}$ values into 200$\times$200 meshes. 

\begin{figure*}
\includegraphics[width=6.8in]{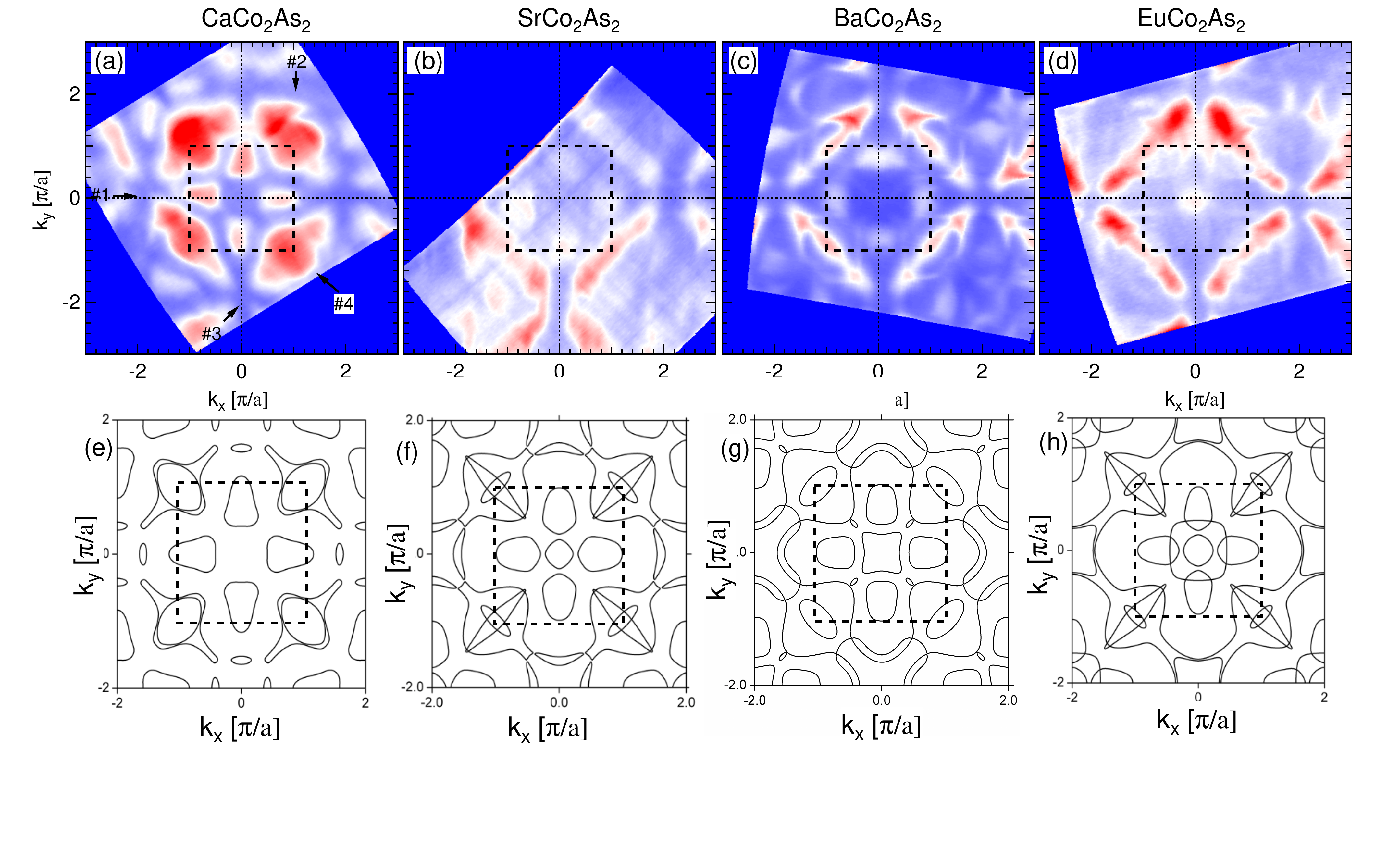}
\caption{Fermi surface maps of {\it A}Co$_2$As$_2$ ({\it A} = Ca, Sr, Ba, Eu) plotted by integrating the photoelectron intensity within $\pm 10$~meV about {\it E}$_{\rm F}$, for (a) CaCo$_2$As$_2$ ($T_{\rm S}\approx20$~K, $h\nu=135$~eV), (b) SrCo$_2$As$_2$ ($T_{\rm S}\approx90$~K, $h\nu=140$~eV), (c) BaCo$_2$As$_2$ ($T_{\rm S}\approx200$~K, $h\nu=135$~eV), and (d) EuCo$_2$As$_2$ ($T_{\rm S}\approx200$~K, $h\nu=135$~eV). The outline of the first Brillouin zone is shown by a black dashed square. The panels (e--h) are same as (a--d), but calculated. The panels (b, c, f, g) are reconstructed from our earlier publications \cite{DhakaPRB13, Pandey13} for comparison within the $A$Co$_2$As$_2$ family.}
\label{fig1}
\end{figure*}

\section{Results and discussion}

In order to understand how the low-energy band structure changes between different members of $A$Co$_2$As$_2$ family, in Figs.~1(a--d) we present the photoemission intensity maps of {\it A}Co$_2$As$_2$ ({\it A} = Ca, Sr, Ba, Eu) at {\it E}$_{\rm F}$ measured at {\it k}$_{\rm z}\approx2\pi/c$. To plot the FS maps, we have integrated the intensity of the photoelectrons within $\pm 10$~meV about $E_{\rm F}$. Interestingly, we find that the shape of the FSs of {\it A}Co$_2$As$_2$ is more complicated \cite{DhakaPRB13, XuPRX13} than those of the parent $A$Fe$_2$As$_2$ compounds of the 122-family \cite{DhakaPRL11}. In particular, the FS map of CaCo$_2$As$_2$ [Fig.~1(a)] clearly exhibits small elliptical pockets centered around the corner ($X$ point) of the BZ. Additionally, we also observe four small pockets around the center ($\Gamma$ point) of the BZ. In the case of SrCo$_2$As$_2$ [Fig.~1(b)], the FS topology is slightly different in the sense that the elliptical pockets at the $X$ point become straight and the four small pockets around the $\Gamma$ point are now less visible. The FS of BaCo$_2$As$_2$ in Fig.~1(c) is significantly different where the pockets at $X$ point break into small segments, and the four small pockets around the $\Gamma$ point are become smaller in size and change in shape. More interestingly, we observe a small circular pocket at the $\Gamma$ point in the FS map of EuCo$_2$As$_2$. At the corner ($X$ point) of the BZ, long straight segments of intensity pocket for the $A=$ Sr, Ba, and Eu samples are clearly different in shape as compared to CaCo$_2$As$_2$. In the case of $A=$ Eu, the intensity of the four small electron pockets (along the $k_x$, $k_y$ directions) around the center of the BZ almost disappeared as compared to CaCo$_2$As$_2$, see Figs.~1(a, d). 

Interestingly, we observe a significant change in the FS topology of $A$Co$_2$As$_2$ family by changing the element at $A$ site (for example in the present study $A=$ Ca, Sr, Ba, and Eu) \cite{DhakaPRB13, XuPRX13, Pandey13}, as compared to the other related parent compounds $A$Fe$_2$As$_2$ \cite{DhakaPRL11, DhakaPRBrc14}. Moreover, the appearance of a small intensity pocket at the $\Gamma$(0,0) point at $E_{\rm F}$ in EuCo$_2$As$_2$ [see Fig.~1(d)] clearly suggests a shift in the Fermi level, which may be different from the other $A$Co$_2$As$_2$ ($A=$ Ca, Sr, Ba) compounds. 

Therefore, to further understand the electronic structure of these samples, we performed theoretical calculations for all the samples at the same ${\it k}$$_{\rm z}$ points and compared the calculated FS maps in Figs.~1(e--h), respectively. 
\begin{figure}[h]
\includegraphics[width=2.9in]{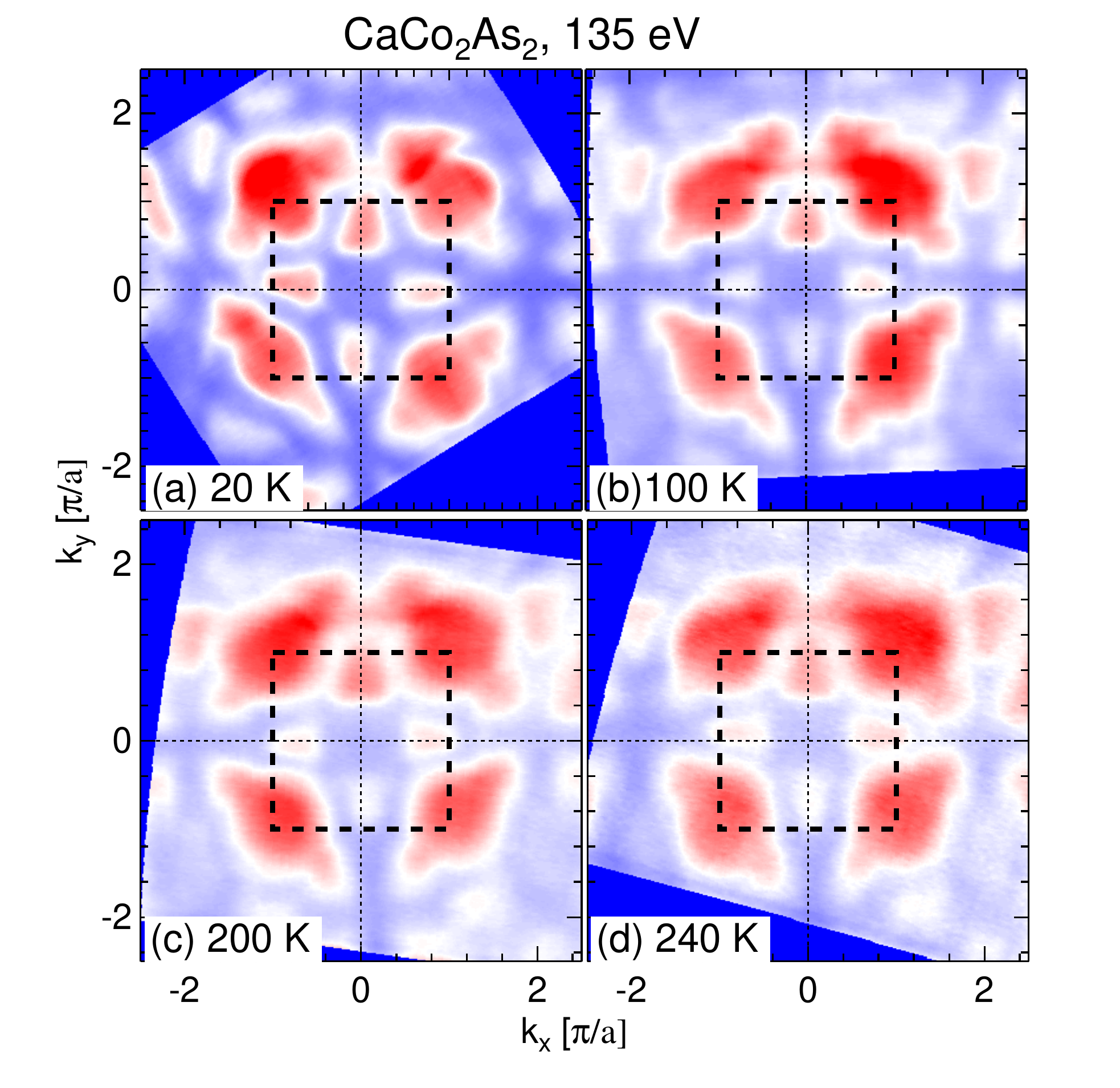}
\caption{Fermi surface maps of CaCo$_2$As$_2$ measured with 135~eV photon energy and at different sample temperatures.}
\label{fig2}
\end{figure}
\begin{figure*}
\includegraphics[width=7.3in]{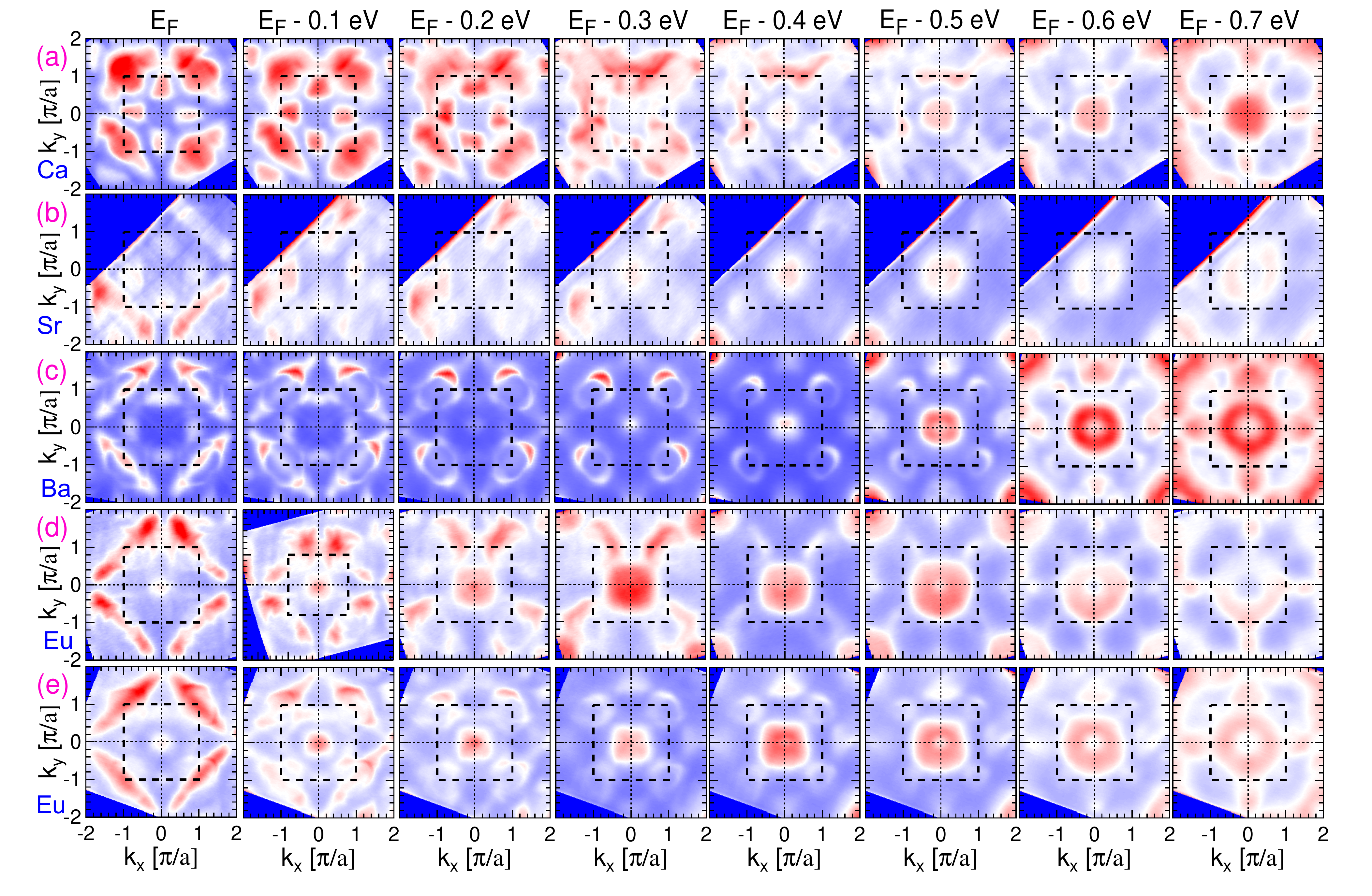}
\caption{The photoemission intensity maps of {\it A}Co$_2$As$_2$ ({\it A} = Ca, Sr, Ba, Eu) plotted at different binding energies, as marked on the top of each panel. Panel (a) CaCo$_2$As$_2$ (135~eV, 20~K), panel (b) for SrCo$_2$As$_2$ (140~eV, 90~K), panel (c) for BaCo$_2$As$_2$ (135~eV, 200~K), and panels (d, e) for EuCo$_2$As$_2$ [(d) 135~eV and (e) 115~eV, 200~K]. The intensity of the photoelectrons is integrated within $\pm 10$~meV about the corresponding binding energy value in each panel.}
\label{fig6}
\end{figure*}
In case of CaCo$_2$As$_2$, four patches of intensity (along the $k_x$, $k_y$ directions) around the $\Gamma$ point and large elliptical pockets across the $X$ points are clearly seen in Fig.~1(e). Overall, the theoretically-calculated FS topology clearly shows significant changes between the Ca, Sr, Ba, and Eu members, see Figs.~1(e--h). Also, the FS maps of {\it A}Co$_2$As$_2$ ({\it A} = Ca, Sr, Ba, Eu) are very different from the other related parent compounds of the 122 family such as CaFe$_2$As$_2$ and BaFe$_2$As$_2$ \cite{DhakaPRBrc14, KondoPRB10}. However, if we see the FS plotted about 400~meV below $E_{\rm F}$ (discussed later) \cite{DhakaPRB13}, it is very similar to that of BaFe$_2$As$_2$, where the FSs  are roughly circular in shape and are roughly similar in size \cite{KondoPRB10}. This clearly indicates a shift of about 400~meV in the Fermi energy by completely replacing Fe with Co in these systems \cite{DhakaPRB13, XuPRX13}. It is important to note here that the overall shape of the measured FSs in Figs.~1(a--d) is in relatively good agreement with the respective calculations in Figs.~1(e--h).  

As discussed above, in case of CaCo$_2$As$_2$ sample, the magnetization measurements show an AFM transition at 52~K \cite{VivekPRB14}. Also, it is important to note here that the ARPES studies of Ba(Fe$_{1-x}$Ru$_x$)$_2$As$_2$ and Ba(Fe$_{1-x}$Co$_x$)$_2$As$_2$ show significant changes in the band structure with sample temperature \cite{DhakaPRL13, Brouet13}. Moreover, it was shown that temperature-dependent FS nesting may play an important role in driving the AFM-paramagnetic phase transition in these materials \cite{LiuPRL09, KondoPRB10, DhakaPRL11, LiuNP10}. Therefore, to get more insights for understanding whether there are changes in the band structure of CaCo$_2$As$_2$ with temperature across the AFM transition temperature ({\it T}$_{\rm N}$), we measured the FSs with 135~eV photon energy using a synchrotron radiation source. In Fig.~2, the FSs are shown at different sample temperatures from 20 to 240~K, which confirm that there are no notable differences in the FS topology measured below [Fig.~2(a)] and above [Figs.~2(b--d)] ($T_{\rm N}$). Further, as the FS of $A$Fe$_2$As$_2$ below $T_{\rm N}$ shows a reconstruction \cite{LiuNP10}, which is clearly absent in the FS of CaCo$_2$As$_2$ measured at 20~K, see Fig.~2(a). These results indicate that the electron correlations are moderate in the $A$Co$_2$As$_2$ family, as the overall mass enhancement of the Co 3$d$ electrons is found to be smaller \cite{MaoPRB18} as compared to the Fe 3$d$ electrons in $A$Fe$_2$As$_2$ based compounds \cite{LiuPRL09, KondoPRB10, DhakaPRL11, LiuNP10}.   

\begin{figure}
\includegraphics[width=3.5in]{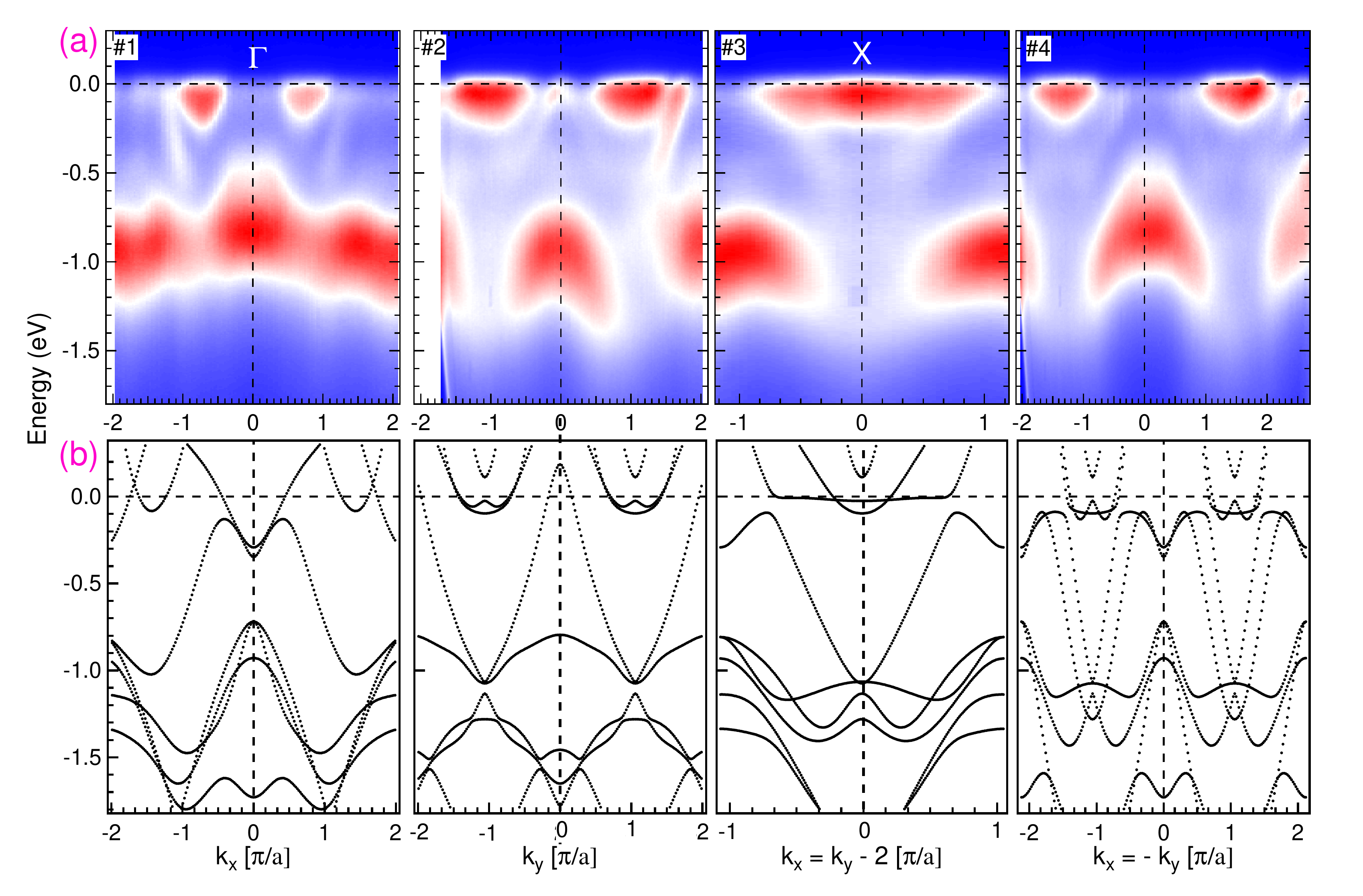}
\includegraphics[width=3.5in]{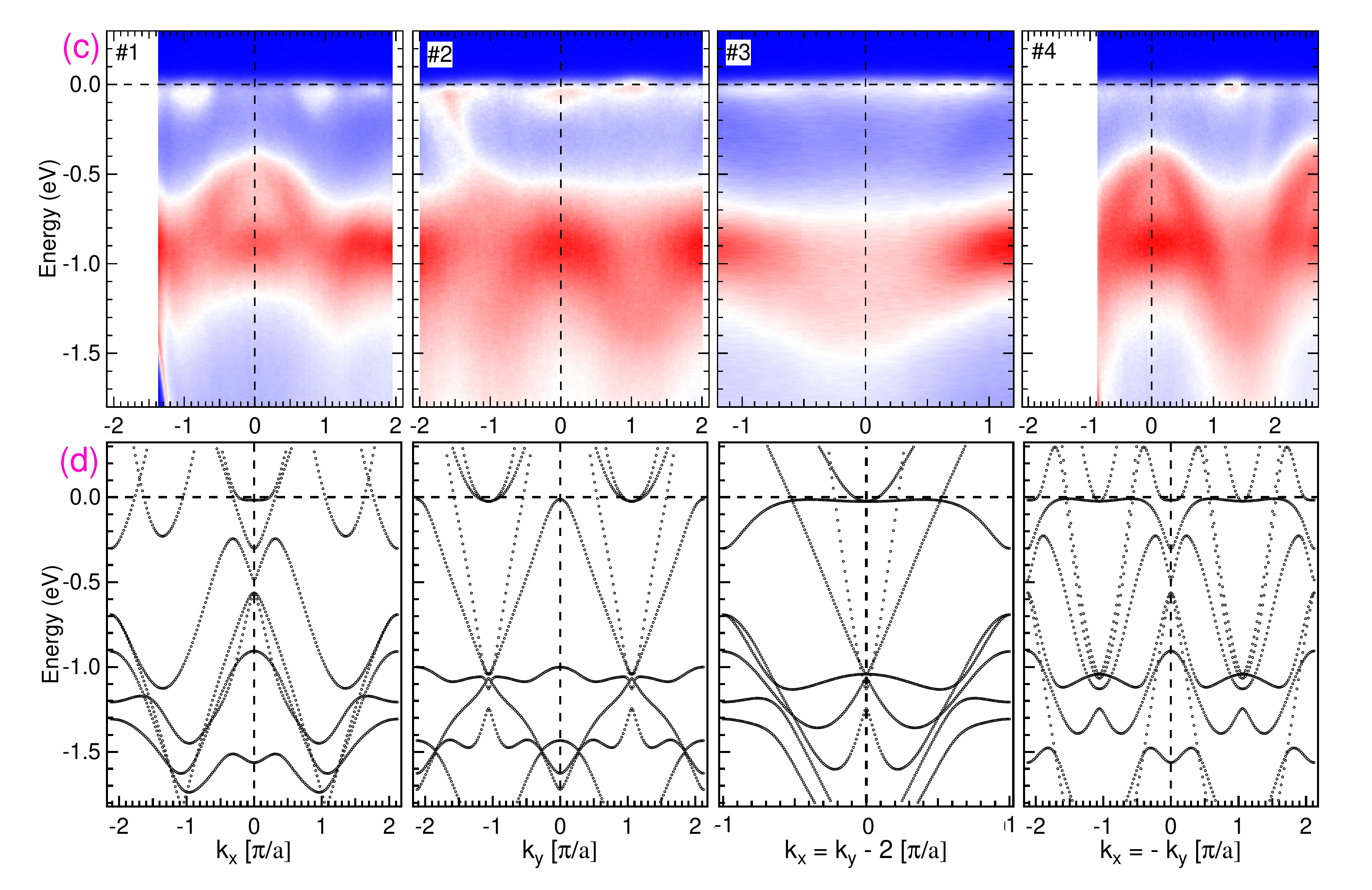}
\caption{Experimental band dispersion data of CaCo$_2$As$_2$ (panel a) and SrCo$_2$As$_2$ (panel c) measured with 135~eV and 140~eV photon energies, and the corresponding theoretical band dispersions in panels (b) and (d), respectively. The locations of the cuts \#1, \#2, \#3 and \#4 are marked in Fig.~1(a). In panels (c) and (d), cuts\#1 and \#3 are similar to the reported in our  publication \cite{Pandey13}, reconstructed for comparison.}
\label{fig4}
\end{figure}
 
In order to compare the FS topology at different binding energies, we show the $(k_{\it x}, k_{\it y})$ plots from 0 to 700~meV below {\it E}$_{\rm F}$ for all the samples in Fig.~3. For the CaCo$_2$As$_2$, we observe a decrease in the intensity of the small pockets around the zone center along with a shape change at the corner of the BZ, see panel (a) in Fig.~3. At 400~meV below {\it E}$_{\rm F}$, a small pocket is clearly seen at the center of the BZ, which grows in size with further increase in the binding energy. At the same time, the shape of the pocket at the corner is very similar to that of the electron pocket observed in $A$Fe$_2$As$_2$ samples. Similar changes are observed for the SrCo$_2$As$_2$ sample, as shown in panel (b) except that the center pocket appeared at around 300~meV below {\it E}$_{\rm F}$, and the size of both the pockets at the center and at the corner of the BZ is almost similar at 400~meV. For the BaCo$_2$As$_2$ sample, a small intensity of the central pocket is visible at 200~meV, see panel (c), and again the size of both pockets is similar at 300-400~meV below {\it E}$_{\rm F}$. Interestingly, in case of the EuCo$_2$As$_2$, the central pocket is already visible at {\it E}$_{\rm F}$, see panels (d, e), measured at 135~eV and 115~eV photon energies, respectively. Further, the size and shape of both pockets at the center and corner of the BZ look similar for all the samples at around 400~meV below {\it E}$_{\rm F}$ except for EuCo$_2$As$_2$. More carefully, if we compare the size of the pocket at the BZ center at 700~meV below {\it E}$_{\rm F}$ for all the samples, it qualitatively increases from Ca to Sr to Ba to Eu. These $(k_{\it x}, k_{\it y})$ plots of $A$Co$_2$As$_2$ further motivated us to investigate the low-energy band structure in detail by plotting the band dispersion and comparing them with those calculated theoretically along different momentum $(k_{\it x}, k_{\it y})$ directions.     

\begin{figure}
\includegraphics[width=3.5in]{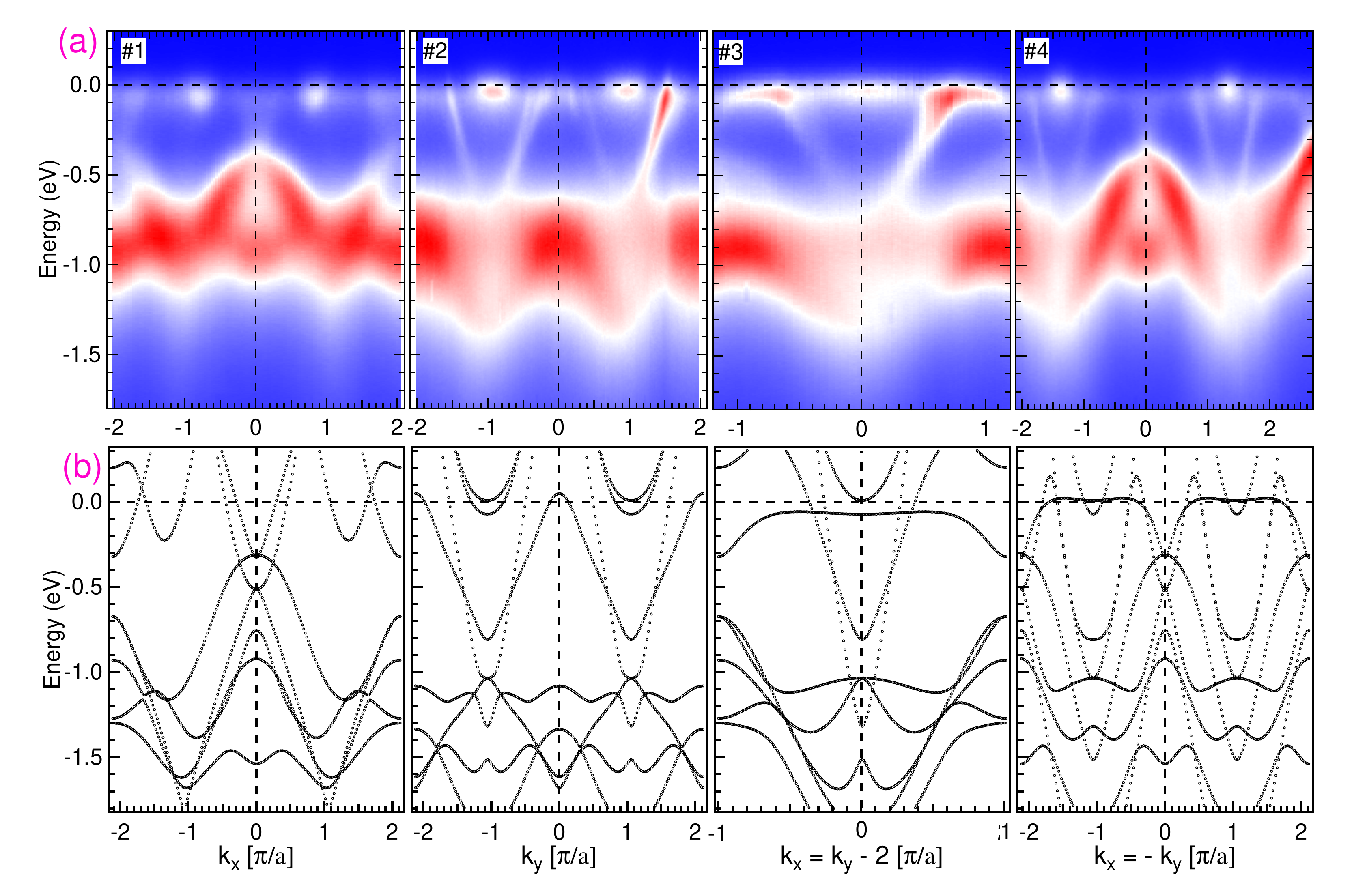}
\includegraphics[width=3.5in]{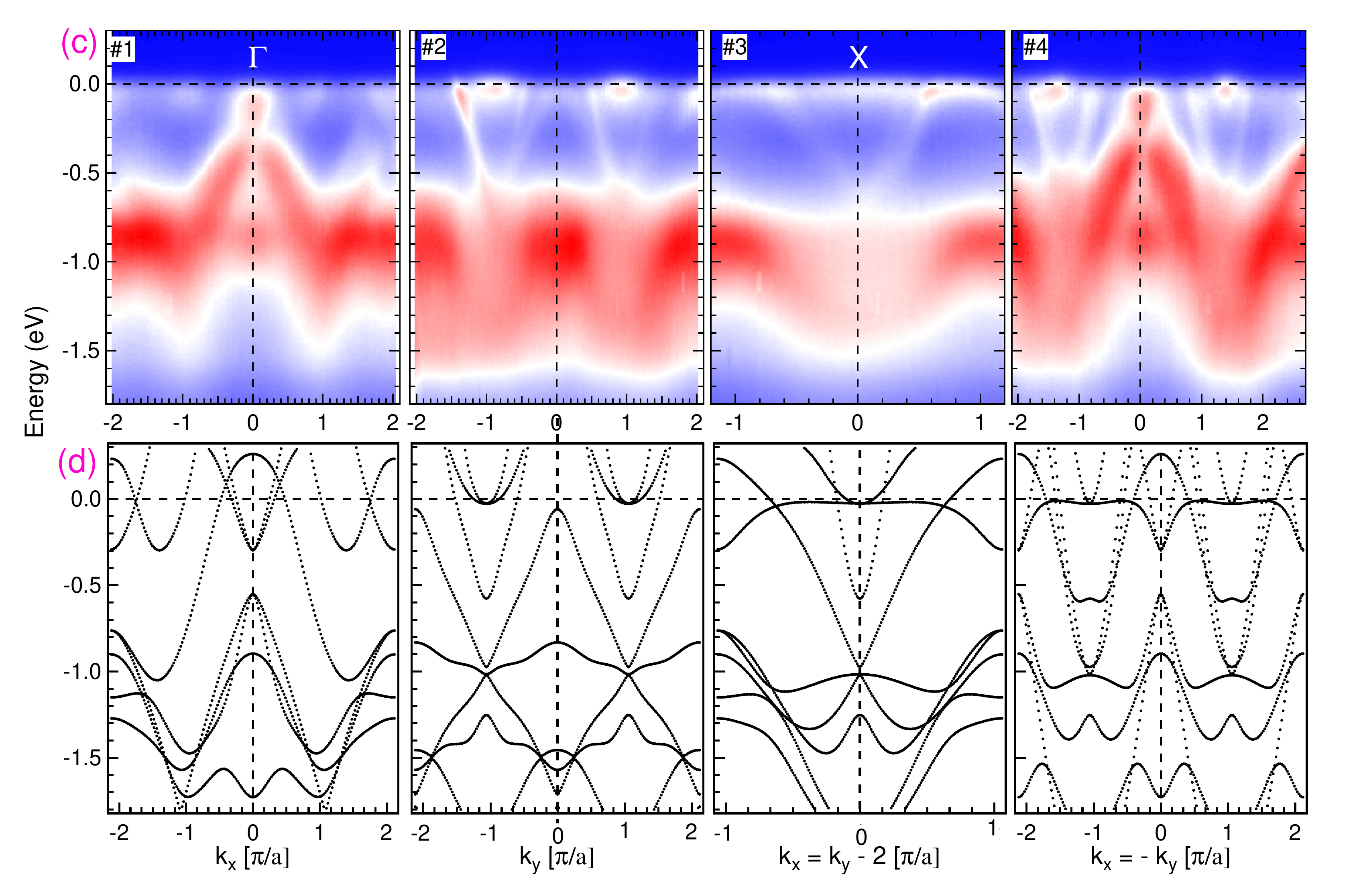}
\caption{Experimental band dispersion data of BaCo$_2$As$_2$ (panel $a$) and EuCo$_2$As$_2$ (panel $c$) measured with 135~eV photon energy. The plots in panels $b$ and $d$ are the corresponding theoretical band dispersions. The locations of the cuts \#1, \#2, \#3 and \#4 are marked in Fig.~1(a).}
\label{fig5}
\end{figure}

Therefore, to elucidate the character of these FS pockets, we extracted in-plane dispersions for all the samples and shown the energy-momentum intensity plots in Figs.~4 and 5 along the four different cuts \#1, \#2, \#3 and \#4. These cuts are from $(k_{\it x}, k_{\it y})=$ ($-$2$\pi$, 0 to 2$\pi$, 0), (2$\pi$, $\pi$ to $-$2$\pi$, $-\pi$), (0, $-$2$\pi$ to 2$\pi$, 0), ($-$2$\pi$, 2$\pi$ to 2$\pi$, $-$2$\pi$), respectively, as marked in Fig.~1(a). The band dispersions measured with 135~eV photon energy are shown in Fig.~4 for the $A$Co$_2$As$_2$ ($A=$ Ca, Sr) samples. The photoelectron intensity along the cut \#1 at the center of the BZ ($\Gamma$ point) is not very clear for the these samples; however, we clearly observe two small electron pockets centered at the Fermi momenta of $k_{\rm F\it x}\approx\pm 0.7 \pi/a$. Interestingly, the top of the hole pocket at the center of the BZ is clearly seen at about 400~meV below $E_{\rm F}$. Moreover, we find a large electron pocket with $k_{\rm F\it y}=\pm 0.65 \pi/a$ in the band dispersion plotted along the cut \#2 at the corner of the first BZ at ${\it X}=(\pi,\pi)$. In the cut \#3, the data show a large electron pocket with $k_{\rm F}=\pm 0.65 \pi/a$ and high intensity peak at the $X$ point. For comparison, we have also plotted the data along the cut \#4, which again shows large electron pockets at the $X$ points. To get more insights, it is important to compare the experimental results with theoretical calculations, and therefore, we present the calculated band dispersions in the lower panels for Fig.~4 along all four different cuts. We find a reasonable qualitative agreement between the theoretical and measured data. 

Moreover, in Fig.~5, we show the low-energy band structure of BaCo$_2$As$_2$ and EuCo$_2$As$_2$ samples to understand the character of the FS pockets. We note here that the band dispersions of BaCo$_2$As$_2$ are very similar to those of SrCo$_2$As$_2$ except that the small-intensity electron-like bands are visible at the center of the BZ for BaCo$_2$As$_2$, which are found to be slightly stronger when measured with $h\nu=$ 85~eV in Ref.~\cite{DhakaPRB13}. Also, the slope of the hole bands is more slanting in $A$Co$_2$As$_2$ ($A=$ Sr, Ba) in comparison to CaCo$_2$As$_2$, see for example cuts \# 1 and \#4 in Figs.~4 and 5. 

It is intriguing to see the band structure of EuCo$_2$As$_2$, which is found to be significantly different from the other $A$Co$_2$As$_2$ ($A=$ Ca, Sr, Ba) samples, see Figs.~4 and 5. We observe that the intensity of two smaller electron pockets at Fermi momenta $k_{\rm F\it x} = \pm 0.9 \pi/a$ reduced significantly and the top of the hole band at the center of the BZ ($\Gamma$ point) is about 300~meV below $E_{\rm F}$, see along the cut \#1. This indicates that in the case of EuCo$_2$As$_2$, the rigid band shift at the center of the BZ is smaller ($\approx$300~meV) than in the other $A$Co$_2$As$_2$ ($A=$ Ca, Sr, Ba) compounds (400~meV). Therefore, the Fermi momentum of the electron pocket at the $\Gamma$ point $k_{\rm F\it x} = \pm 0.15 \pi/a$ is also slightly smaller than in BaCo$_2$As$_2$ \cite{DhakaPRB13}. At the corner of the first BZ [$X=(\pi,\pi)$], we observe a larger electron pocket for $A$Co$_2$As$_2$ with $k_{\rm F}=\pm 0.65 \pi/a$ (along both the cuts \#2 and \#3) as compared to the {\it A}Fe$_2$As$_2$ compounds \cite{DhakaPRL11}. The data through cut \#4 again confirm the presence of large electron pockets at the $X$ points and a smaller electron pocket at the $\Gamma$ point. The band structure plots obtained from the FP-LAPW calculations, shown in the lower panels of Fig.~5, are in reasonable agreement with the experimentally-observed band dispersions. 

Now we discuss the broad comparison of the band structure and FS topology between {\it A}Co$_2$As$_2$ and {\it A}Fe$_2$As$_2$ compounds. Overall, the electronic DOS($E$) of the {\it A}Co$_2$As$_2$ samples are found to be similar to those of the Fe-based compounds. However, we observe a band shift of 300--500~meV below {\it E}$_{\rm F}$ due to extra $d$ electron in Co as compared to Fe \cite{DhakaPRB13, Pandey13, Sefat09, Singh09}, which may reflect the high intensity peak at {\it E}$_{\rm F}$. Note that the Co $d_{x^2-y^2}$ orbital has a larger bandwidth (between $-5$~eV and 2~eV) than the Fe $d$ orbitals (from $-2$~eV to 2~eV) \cite{SinghPRL08}. Therefore, a complex multi-band Fermi surface is observed in the ARPES measurements on the $A$Co$_2$As$_2$ compounds along with the large DOS at {\it E}$_{\rm F}$ with no apparent nesting \cite{DhakaPRB13, XuPRX13}. In this scenario, the band structure of {\it A}Co$_2$As$_2$ \cite{MaoPRB18} appears significantly different at {\it E}$_{\rm F}$ with respect to that of BaFe$_2$As$_2$. Also, as noted above, CaCo$_2$As$_2$ shows magnetic ordering at low temperature; however, APRES measurements across the magnetic transition indicate no significant changes in the band structure. Similarly, a complex FS structure of EuRh$_2$As$_2$ and BaNi$_2$As$_2$ has been reported using ARPES, but no signature of band folding due to magnetic ordering was observed, which indicates a weak coupling between layers \cite{PalczewskiPRB12, Zhou11}. Moreover, a significant decrease in electronic correlation is reported in BaCr$_2$As$_2$ \cite{RichardPRB17, NayakPNAS17}. Note that in the $A$Co$_2$As$_2$ compounds a considerable decrease in interlayer distance and {\it z}$_{\rm As}$ results in different correlation strengths \cite{MaoPRB18} when compared with iron pnictides \cite{YinNM11}. 

\begin{figure}[h]
\includegraphics[width=3.6in]{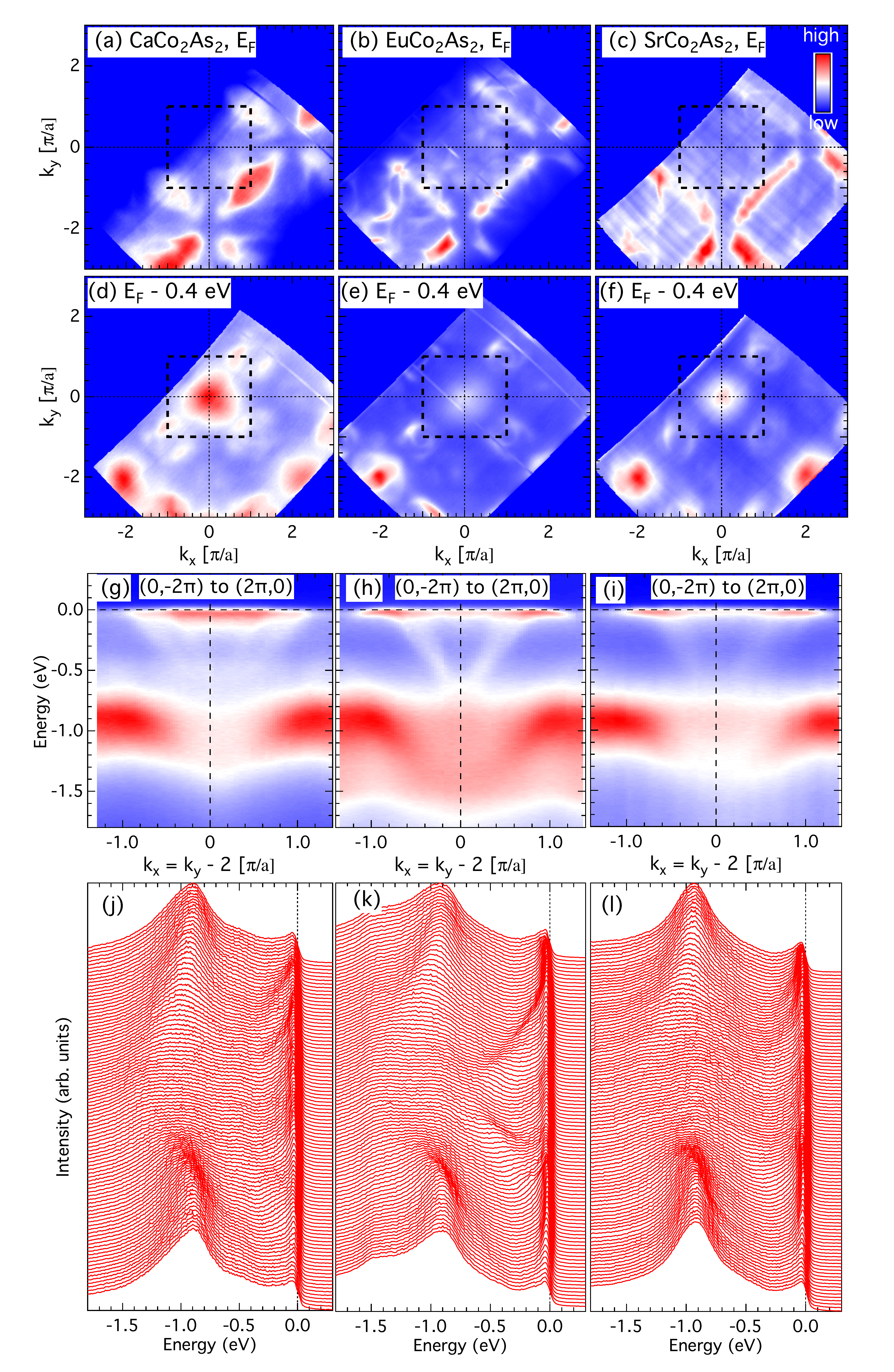}
\caption{(a--c) Fermi surface (FS) maps of $A$Co$_2$As$_2$ ($A=$ Ca, Eu, Sr) measured with photon energy of 100~eV and at 90~K, where the first BZ boundaries are shown by black-dashed square. (d--f) Photoemission intensity maps are plotted at $E_{\rm F}-0.4$~eV. These maps in (a--f) are constructed by integrating the intensity of the photoelectrons within $\pm 10$~meV. (g--i) The band dispersions along the (0,$-$2$\pi$ to 2$\pi$,0) direction from (a--c), respectively, and the corresponding energy dispersive curves (EDCs) are shown in (j--l).}
\label{fig7}
\end{figure} 

Finally, we compare the FSs and low-energy band structures, in Fig.~6 of $A$Co$_2$As$_2$ ($A=$ Ca, Eu, Sr) measured at 90~K with 100~eV photon energy. The intensity of photoelectrons along ($k_{\it x}$, $k_{\it y}$) is plotted at $E_{\rm F}$ in Figs.~6(a--c) and at 400~meV below $E_{\rm F}$ in Figs.~6(d--f). The FS plots at $E_{\rm F}$ show a long segment of intensity across the $X$ points (corners of the BZ). The FS plots at 400~meV below $E_{\rm F}$ show clear changes in the FS shape at the $X$ point, which changes from a large segment to an oval--shape pocket and approximately similar size as the pocket at the center of the BZ ($\Gamma$ point). These results demonstrate the rigid band-like shift of 300-400~meV (depending on $A$) below $E_{\rm F}$ for $A$Co$_2$As$_2$ as compared to $A$Fe$_2$As$_2$. Figures~6(g--i) show the band dispersion data plotted across the corner of the BZ ($X$ point) i.e. from (0, $-$2$\pi$) to (2$\pi$, 0) and the corresponding energy dispersive curves (EDCs) are shown in Figs.~6(j--l). Again, for all these compounds we observe a large electron pocket at the $X$ point and the bottom of this pocket is about 500~meV below $E_{\rm F}$. More interestingly, we reveal a rather large flat band, driven by the Co 3$d_{x^2-y^2}$ orbital, in close vicinity of $E_{\rm F}$ with strong intensity near the $X$ point in both the experimental and calculated data of $A$Co$_2$As$_2$ compounds \cite{DhakaPRB13}, as shown in Fig.~4 [cuts \#2, \#3 and \#4 of panel (a)] as well as in Fig.~6. In ARPES measurements on the Fe$_{1.03}$Te$_{0.94}$S$_{0.06}$ superconductor, Starowicz $et al.$ observed a flat band at the Fermi level, which gives rise to high density of states at $E_{\rm F}$ \cite{Starowicz13}. This was explained in terms of as a Van Hove singularity, which is believed to play an important role in the emergence of superconductivity in Fe-based compounds \cite{Starowicz13}. The presence of a large density of states near $E_{\rm F}$ due to a nearly flat band at the corner of the BZ is crucial to understand the physical properties of the $A$Co$_2$As$_2$ compounds.  

\section{Conclusions}

We have presented a comprehensive study of the electronic properties of the $A$Co$_2$As$_2$ ($A=$ Ca, Sr, Ba, and Eu) compounds using ARPES and theoretical FP-LAPW calculations. The FSs of these compounds are different from those of the parent compounds of FeAs-based high-temperature superconductors. The band dispersion data show a small electron pocket at the center and large electron pockets at the corner of the Brillouin zone. The experimental data agree reasonably well with the theoretical calculations. The absence of the FS nesting in $A$Co$_2$As$_2$ is in contrast to the $A$Fe$_2$As$_2$ compounds. However, the top of the hole bands is found to be moved 300--400~meV below {\it E}$_{\rm F}$ (depending on $A$) resulting an appearance of a small electron pocket at the center of the BZ. More interestingly, we observe large flat bands near {\it E}$_{\rm F}$, which result in a large density of states and could be responsible for the interesting physical properties of these materials. Furthermore, no significant changes are observed in the FS topology of CaCo$_2$As$_2$ between 20 and 300~K across the AFM transition. We discuss similarities and differences in the electronic properties of $A$Co$_2$As$_2$ with respect to the parent {\it A}Fe$_2$As$_2$ compounds of 122 family.\\

\section{Acknowledgments}

We thank Aaron Bostwick and Eli Rotenberg for excellent support at the ALS.  This research was supported by the U.S. Department of Energy, Office of Basic Energy Sciences, Division of Materials Sciences and Engineering. Ames Laboratory is operated for the U.S. Department of Energy by Iowa State University under Contract No. DE-AC02-07CH11358. The Advanced Light Source is supported by the Office of Basic Energy Sciences, U. S. Department of Energy under Contract No. DE-AC02-05CH11231. RSD also acknowledges the support by BRNS through a DAE Young Scientist Research Award with Project Sanction No. 34/20/12/2015/BRNS.

\end{document}